\documentclass[preprints,article,accept,pdftex,moreauthors]{Definitions/mdpi} 
\firstpage{1} 
\pubvolume{1}
\issuenum{1}
\articlenumber{0}
\pubyear{2023}
\copyrightyear{2023}
\datereceived{ } 
\daterevised{ } 
\dateaccepted{ } 
\datepublished{ } 
\hreflink{https://doi.org/} 
\pdfoutput=1 

\Title{Hydrodynamic Simulations of a Relativistic Jet Interacting with the Intracluster Medium: Application to Cygnus A}

\TitleCitation{Title}

\newcommand{\code}[1]{\texttt{#1}}

\newcommand\chandra{\textit{Chandra}}

\newcommand\ic{inverse Compton}

\Author{John A. ZuHone $^{1,*}$\orcidA{}, Paul E. J. Nulsen $^{1,2}$, Po-Hsun Tseng $^{3}$, Hsi-Yu Schive
$^{3,4}$ and Tom W. Jones $^{5}$}

\AuthorNames{John ZuHone, Paul Nulsen, Po-Hsun Tseng, Hsi-Yu Schive and Tom Jones}

\AuthorCitation{ZuHone, J.; Nulsen, P.; Tseng, P.-H.; Schive, H.-Y.; Jones, T.}

\address{%
$^{1}$ \quad Center for Astrophysics $\vert$~Harvard \& Smithsonian, 60 Garden St., Cambridge, MA 02138, USA\\
$^{2}$ \quad ICRAR, University of Western Australia, 35 Stirling Hwy, Crawley, WA 6009, Australia\\
$^{3}$ \quad Institute of Astrophysics, National Taiwan University, Taipei 10617, Taiwan\\
$^{4}$ \quad Physics Division, National Center for Theoretical Sciences, Taipei 10617, Taiwan\\
$^{5}$ \quad School of Physics and Astronomy, University of Minnesota Twin Cities, Minneapolis, MN, USA}

\corres{Correspondence: john.zuhone@cfa.harvard.edu}

\abstract{The Fanaroff-Riley class II radio galaxy Cygnus A hosts jets which produce radio emission, X-ray cavities, cocoon shocks, and X-ray hotspots where the jet interacts with the ICM. Surrounding one hotspot is a peculiar ``hole'' feature which appears as a deficit in X-ray emission. We use relativistic hydrodynamic simulations of a collimated jet interacting with an inclined interface between lobe and cluster plasma to model the basic processes which may lead to such a feature. We find that the jet reflects off of the interface into a broad, turbulent flow back out into the lobe, which is dominated by gas stripped from the interface at first and from the intracluster medium itself at later times. We produce simple models of X-ray emission from the ICM, the hotspot, and the reflected jet to show that a hole of emission surrounding the hotspot as seen in Cygnus A may be produced by Doppler de-boosting of the emission from the reflected jet as seen by an observer with a sight line nearly along the axis of the outgoing material.
}


\begin{document}


\section{Introduction}

Due to its proximity and power, the Fanaroff-Riley class II radio galaxy (FRII) \citep{fr74}, Cygnus A, has served as the archetype of powerful radio galaxies \citep{cb96}.  With its remarkable radio properties, Cygnus A is hosted by the central galaxy of a massive cluster, where the radio AGN has been pumping about $10^{46}\rm\ erg\ s^{-1}$ through jets into the intracluster medium (ICM) over the past $\simeq 20$ Myr \citep{gs13,snw18}.  Demonstrating the impact a radio galaxy can have on its environment, Cygnus A provides a prime example of radio mode AGN feedback \citep{csw06,mn07,f12}

\begin{figure}
\centering
\includegraphics[width=0.95\textwidth]{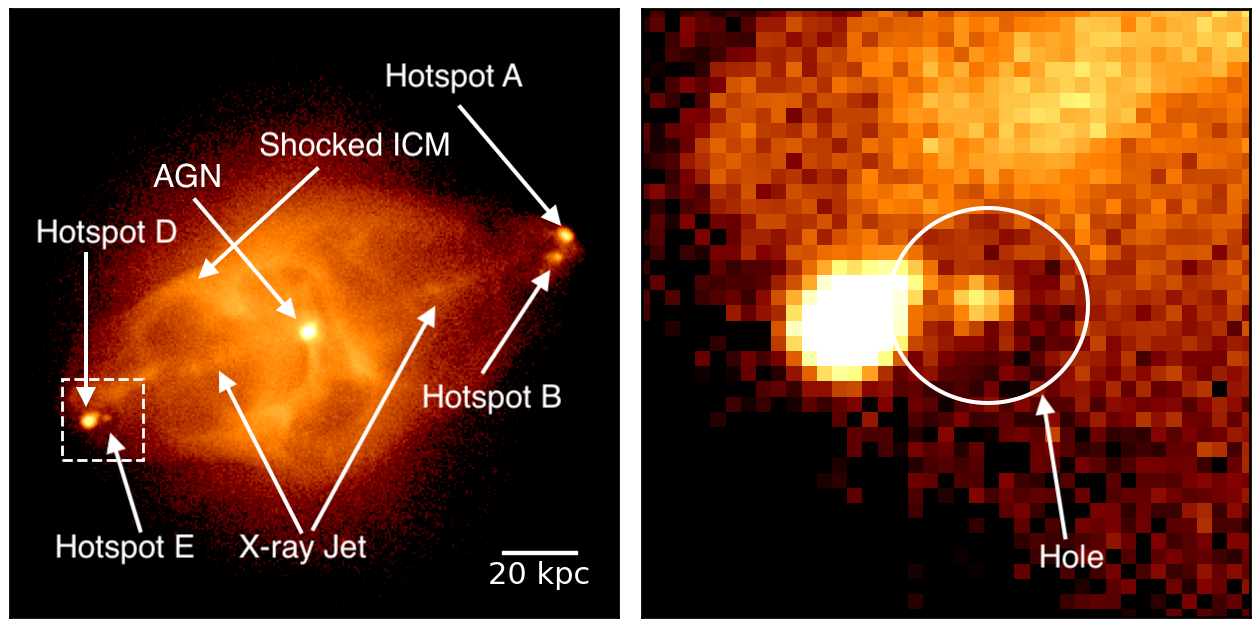}
\caption{Exposure-corrected \chandra{} observation of Cygnus A in the 0.5-7~keV band. Left panel: Entire region containing the shocked ICM, hotspots, X-ray jet, and AGN. Right panel: Zoom-in (shown in the left panel) on hotspots D and E, showing in more detail the hole of emission surrounding hotspot E. The contrast in the right panel has been altered for clarity. Reproduced from \cite{sjn20}.\label{fig:cyga}}
\end{figure}

Deep \chandra{} X-ray observations have been used to probe the impacts of Cygnus A on its surroundings, revealing complex structure created by the radio outburst.  This includes X-ray cavities, cocoon shocks and X-ray hotspots, which are found in a number of other cluster central radio galaxies \citep{mn07,Gitti2012,Hardcastle2020}.  Cygnus A also has a unique ``X-ray jet,'' thought to be \ic{} radiation scattered by energetic particles that are relics of the jets accumulated over time \citep{sbd08}. These features are shown in the \chandra{} image of Cygnus A in Figure \ref{fig:cyga}, left panel. Regions of more diffuse enhanced X-ray emission from its lobes are well correlated with enhancements in radio emission, supporting the case that relativistic electrons spread throughout the lobes also produce significant levels of more diffuse \ic{} X-ray emission \citep{chh05,dwh18,sjn20}.  

Another novel feature revealed by the deep \chandra{} observations of Cygnus A is a roughly circular X-ray "hole," $\simeq 4$ kpc in radius, in the region surrounding hotspot E in the eastern lobe \citep{sjn20} (see Figure \ref{fig:cyga}, right panel). To account for the X-ray deficit over the hole, the \ic{} emission seen in the rest of the lobe must be reduced or absent from a region elongated along our line of sight, with a depth exceeding its diameter by at least a factor of $\simeq 1.7$ \citep{sjn20}. Assuming that the hotspot is formed where the jet strikes the ICM \citep{br74}, the implication is that, after striking the shocked ICM, jet plasma flows back into the lobe, displacing relativistic plasma from the hole region.  The jet is likely to have encountered one or more strong shocks in the hotspot, causing a significant population of electrons to be accelerated to relativistic energies, as required to produce the observed synchrotron-self Compton X-ray emission from the hotspot \citep{hcp94}. However, as pointed out originally by \cite{sjn20}, if the outflow from the hotspot is directed away from the earth at sufficient speed, Doppler beaming can reduce the \ic{} X-ray emission directed towards the earth enough to create the apparent hole. Such Doppler de-boosting would similarly produce a hole in the radio band, which is also observed \cite{sjn20}.

This paper discusses simulations of the encounter between the jet and ICM at hotspot E in Cygnus A, in an effort to test the Doppler de-boosting hypothesis for the presence of the hole suggested by \cite{sjn20}. A hydrodynamic code is employed to simulate an unmagnetized, relativistic jet flowing through a low density lobe before it encounters the inclined interface between the lobe and the far denser ICM.  Details of the simulations are described in section \ref{sec:methods}, including the physical model of the fluid (Section \ref{sec:physics}), the numerical code (Section \ref{sec:code}), and the setup of the flow model (Section \ref{sec:setup}). The analysis of the results is presented in Section \ref{sec:results}, and we present and discuss our conclusions in Section \ref{sec:summary}.

\section{Methods}\label{sec:methods}

\subsection{Physics}\label{sec:physics} 

The mass, momentum, and energy conservation laws of a special relativistic fluid
are 
\begin{eqnarray}
\partial_\nu(\rho{U}^\nu) &=& 0, \label{eqn:mass_cons} \\
\partial_\nu{T^{\mu\nu}} &=& 0, \label{eqn:se_cons}
\end{eqnarray}
where the Einstein summation convention is employed over repeated indices here and throughout this work.
The stress-energy tensor for an ideal relativistic fluid is 
\begin{equation}\label{eqn:stress_energy}
T^{\mu\nu} = \rho{h}{U^\mu}{U^\nu}/c^2 + p\eta^{\mu\nu}.
\end{equation}
In these equations, $\rho$ and $p$ are the proper mass density and the pressure,
$U^\nu$ is the four-velocity, $\eta^{\mu\nu}$ is the metric tensor (signature $-, +, +, +$) for Minkowski
spacetime, $c$ is the speed of light, and $h$ is the specific enthalpy, given by
\begin{equation}\label{eqn:enthalpy}
h = c^2+\epsilon+\frac{p}{\rho}
\end{equation}
where $\epsilon$ is the specific thermal energy. Note that although the \code{GAMER}  code \citep{Tseng2021} that we use for our simulation runs carries out calculations in units where $c = 1$, we include the
factors of $c$ here for completeness.

This set of equations is closed by an equation of state (EOS) $h(\rho,p)$; the EOS that we employ in
this work is the Taub-Mathews (TM) EOS \cite{Taub1948,Mathews1971,Mignone2005}, which is an
approximation to the Synge model, the exact EOS 
for an ideal, non-degenerate gas composed of a single particle species 
\cite{Synge1958}.  The TM EOS is given by
\begin{equation}\label{eqn:tm_eos}
\frac{h_{\rm TM}}{c^2} = 2.5\left(\frac{k_BT}{mc^2}\right) + \sqrt{2.25\left(\frac{k_BT}{mc^2}\right)^2 + 1}
\end{equation}
where $k_B$ is the Boltzmann constant, $T$ is the gas temperature, and $m$ is the mass per particle
in the fluid. The use of the TM EOS enables the relativistic and non-relativistic gas phases to be combined in a single, realistic simulation, which cannot be achieved with a standard polytropic
EOS with a single value for the ratio of specific heats $\Gamma$. For the TM EOS, $\Gamma \rightarrow 5/3$ for a
non-relativistic fluid and $\Gamma \rightarrow 4/3$ for a relativistic fluid, and the transition
between the phases occurs near $k_BT \sim mc^2$ (see Figure 1 of \cite{Tseng2021}). 

\subsection{Code}\label{sec:code}

To perform our simulation runs we use \code{GAMER} \cite{Schive2018}, an
adaptive-mesh refinement (AMR) astrophysical hydrodynamic code which has a
module for special relativistic hydrodynamics (SRHD) \cite{Tseng2021}. Since we are operating in a regime where both relativistic and non-relativistic gases are present in the same simulation, there is the risk of catastrophic cancellation in a number of expressions when evolving the numerical versions of Equations \ref{eqn:mass_cons}-\ref{eqn:stress_energy}. \code{GAMER} avoids these cancellations by evolving the equation for the reduced energy density separately and computing Lorentz factors using the four-velocity $U^\nu$ rather than the three-velocity $v_i$. For more details about the SRHD solver in \code{GAMER}, we refer the reader to Section 2 of
\cite{Tseng2021}. In our simulations, we use the Harten-Lax-van Leer with Contact (HLLC) solver \cite{Toro1994} to solve the Riemann problem and employ the Piecewise Parabolic Method (PPM) \cite{Colella1984} for state estimation.

\subsection{Simulation Setup}\label{sec:setup}

Our simulation setup consists of a cubical domain 20~kpc on a side (with coordinates in the range
[-10, 10]~kpc), divided between two gas phases in pressure equilibrium, ``ICM'' (the thermal
plasma) and ``lobe'' (the relativistic plasma surrounding the X-ray jet). The interface
between the two phases is a plane extending from end to end along the $z$-axis. To test the
specific hypothesis that the jet reflecting off of the shocked ICM surrounding the lobe produces the
hole, we assume that the interface is inclined with respect to the trajectory of the jet. A jet
hitting the ICM at a right angle would slowly bore into it without reflection.This inclination is
determined by an angle $\theta$ from the $y$-axis of the simulation box. For all of the
simulations, the plane of the interface passes through the point $(x_0, y_0) = (-7, 0)$~kpc. The
interface between the ICM and lobe phases is not abrupt but the density $\rho(x')$ continuously
transitions from one to the other using the functional form:
\begin{eqnarray}
x' &=& (x-x_0)\cos{\theta} - (y-y_0)\sin{\theta} \\
\rho(x') &=& \frac{\rho_{\rm ICM,0}+\rho_{\rm lobe,0}e^{x'/w}}{1+e^{x'/w}}
\end{eqnarray}
where $x'$ is the distance perpendicular to the interface, $\rho_{\rm ICM,0}$ and $\rho_{\rm lobe,0}$
are the densities of the ICM and lobe far away from the interface, respectively, and $w$ is a parameter controlling the width of
the interface in the direction perpendicular to the plane separating the two phases. The parameters for the interface for the ``fiducial'' setup are given in
Table~\ref{tab:sim_params}. The lobe phase is 10$^6$ times less dense and 10$^6$ times hotter than
the ICM phase, such that its gas is a fully relativistic plasma with $\Gamma = 4/3$, whereas the ICM
is non-relativistic with $\Gamma = 5/3$. 

\begin{table}
\caption{Parameters for the ``fiducial'' simulation.\label{tab:sim_params}}
\centering
\begin{tabular}{ll}
\hline
Parameter & Value \\
\hline
$\rho_{\rm ICM,0}$ & $5.66 \times 10^{-26}$~g~cm$^{-3}$ \\
$kT_{\rm ICM,0}$ & 9.52~keV \\
$\rho_{\rm lobe,0}$ & $5.66 \times 10^{-32}$~g~cm$^{-3}$ \\
$p$ & $8.6 \times 10^{-10}$~erg~cm$^{-3}$ \\
$w$ & 10~pc \\
$\rho_{\rm jet,0}$ & $5.66 \times 10^{-31}$~g~cm$^{-3}$ \\
$r_{\rm jet}$ & 0.25~kpc \\
$P_{\rm jet}$ & $4.427 \times 10^{45}$~erg~s$^{-1}$ \\
$\theta$ & 25$^\circ$ \\
$U_s$ & -4.482$c$ \\
$U_0$ & 7.342$c$ \\
\hline
\end{tabular}
\vspace{-1mm}
\end{table}

\begin{figure}
\centering
\includegraphics[width=0.48\textwidth]{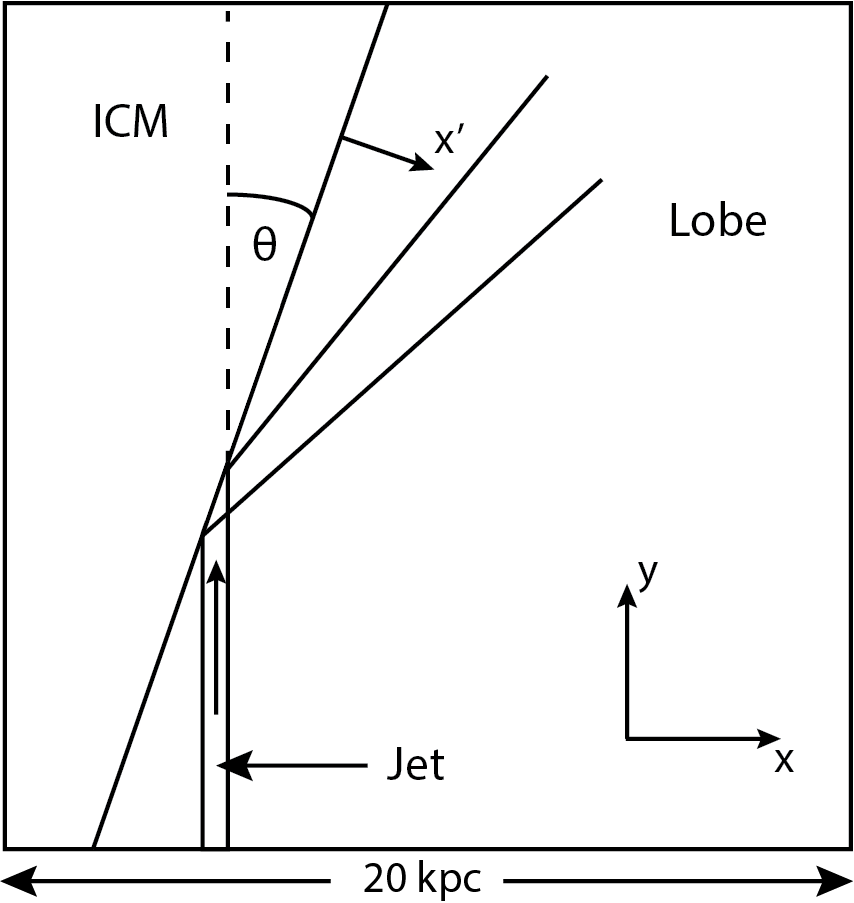}
\caption{Schematic of the simulation setup, showing the ICM and lobe regions 
divided by a sharp interface tilted with respect to the coordinate axes, along 
with the trajectory of the jet.\label{fig:jet_schematic}}
\end{figure}

The jet enters the domain at the $-y$ boundary. Its plasma is 10 times more dense than that of the
lobe, and in pressure equilibrium with it. It enters the domain with a radius of $r_{\rm jet}$, with
a velocity entirely in the $+y$-direction. The rest of the $-y$ boundary is set to a ``diode''
boundary condition, where the gradient of physical quantities is zero across the interface, but the
fluid is only permitted to flow outward. The boundary conditions on the other sides of the domain,
in the $\pm{x}$, $+y$, and $\pm{z}$ are ``outflow'', where the gradient of physical quantities is
zero across the interface. Figure \ref{fig:jet_schematic} shows a rough schematic of the geometry of the interface and the jet.

We determine the velocity of the jet using the equation for the jet power $P_{\rm jet}$ by computing the energy flux through an area $dA_i$ (via Equations \ref{eqn:mass_cons}-\ref{eqn:enthalpy}):
\begin{equation}\label{eqn:jet_power}
P_{\rm jet} = \int [(\gamma-1)\rho{c}^2+\gamma(\rho{\epsilon}+p)]U^idA_i 
\end{equation}
where $\gamma = 1/\sqrt{1-\beta^2}$ is the Lorentz factor and $\beta = v/c$. We assume $P_{\rm jet} = 4.427 \times 10^{45}$~erg~s$^{-1}$, based on the observations of \cite{snw18}. Given the values of $\rho_{\rm jet}$, $r_{\rm jet}$, $p$, and assuming $\Gamma = 4/3$ inside the jet, we can derive a constant jet velocity of $v \sim 0.976c$, or $\gamma\beta \sim 4.482$. 

However, some observations of FR-II jets \cite{w09} suggest that the ``spine'' of the jet (central part) is moving faster than its ``sheath'' (outer parts). To achieve this, we simply assume that the velocity profile with cylindrical radius outward from the jet center is linear:
\begin{equation}\label{eqn:velocity_profile}
U^y(r) = U_s\left(\frac{r}{r_{\rm jet}}\right)+U_0
\end{equation}
Using Equations \ref{eqn:jet_power} and \ref{eqn:velocity_profile}, we assume  $U_s = -4.482c$ and find $U_0 = 7.342c$ such that the mass-weighted average $\langle{U^y}\rangle \approx \gamma\beta c = 4.482c$.

The simulations also evolve passive scalar fields $\rho_c$ for each of the four components, $c$: ``ICM'', ``lobe'', ``interface'', and ``jet''.  The first three fields are defined in the following manner at the beginning of the simulation:
\begin{eqnarray}
\rho_{\rm ICM}/\rho &=& 
    \begin{cases}
        \frac{\rho-0.4\rho_{\rm ICM,0}}{0.4\rho_{\rm ICM,0}} & {\rm if}~0.4\rho_{\rm ICM,0} < \rho < 0.8\rho_{\rm ICM,0} \\
        0 & {\rm if}~\rho \leq 0.4\rho_{\rm ICM,0} \\
        1 & {\rm if}~\rho \geq 0.8\rho_{\rm ICM,0}
    \end{cases} \\
\rho_{\rm lobe}/\rho &=& 
    \begin{cases}
        \frac{2.5\rho_{\rm lobe,0}-\rho}{1.25\rho_{\rm lobe,0}} & {\rm if}~1.25\rho_{\rm lobe,0} < \rho < 2.5\rho_{\rm lobe,0} \\
        0 & {\rm if}~\rho \geq 2.5\rho_{\rm lobe,0} \\
        1 & {\rm if}~\rho \leq 1.25\rho_{\rm lobe,0}
    \end{cases} \\
\rho_{\rm int} &=& \rho-\rho_{\rm ICM}-\rho_{\rm lobe}
\end{eqnarray}
whereas the gas continuously injected within the jet has $\rho = \rho_{\rm jet} = \rho_{\rm jet,0}$ at its entry point. These fields are advected along with the gas.  

We refine the adaptive mesh on the gradients of the density $\rho$, pressure $P$, and Lorentz factor
$\gamma$, which ensures that the ICM/lobe interface, the incident jet, and the reflected jet are all
adequately resolved. The base grid has 64 cells on a side, and we employ 5 levels of refinement
above the base grid level, yielding a smallest cell size of $\Delta{x} = 9.765$~pc.

\section{Results}\label{sec:results}

\subsection{Fiducial Run: Slices}\label{sec:fiducial_slices}

\begin{figure*}
\begin{adjustwidth}{-\extralength}{0cm}
\begin{center}
\includegraphics[width=0.49\fulllength]{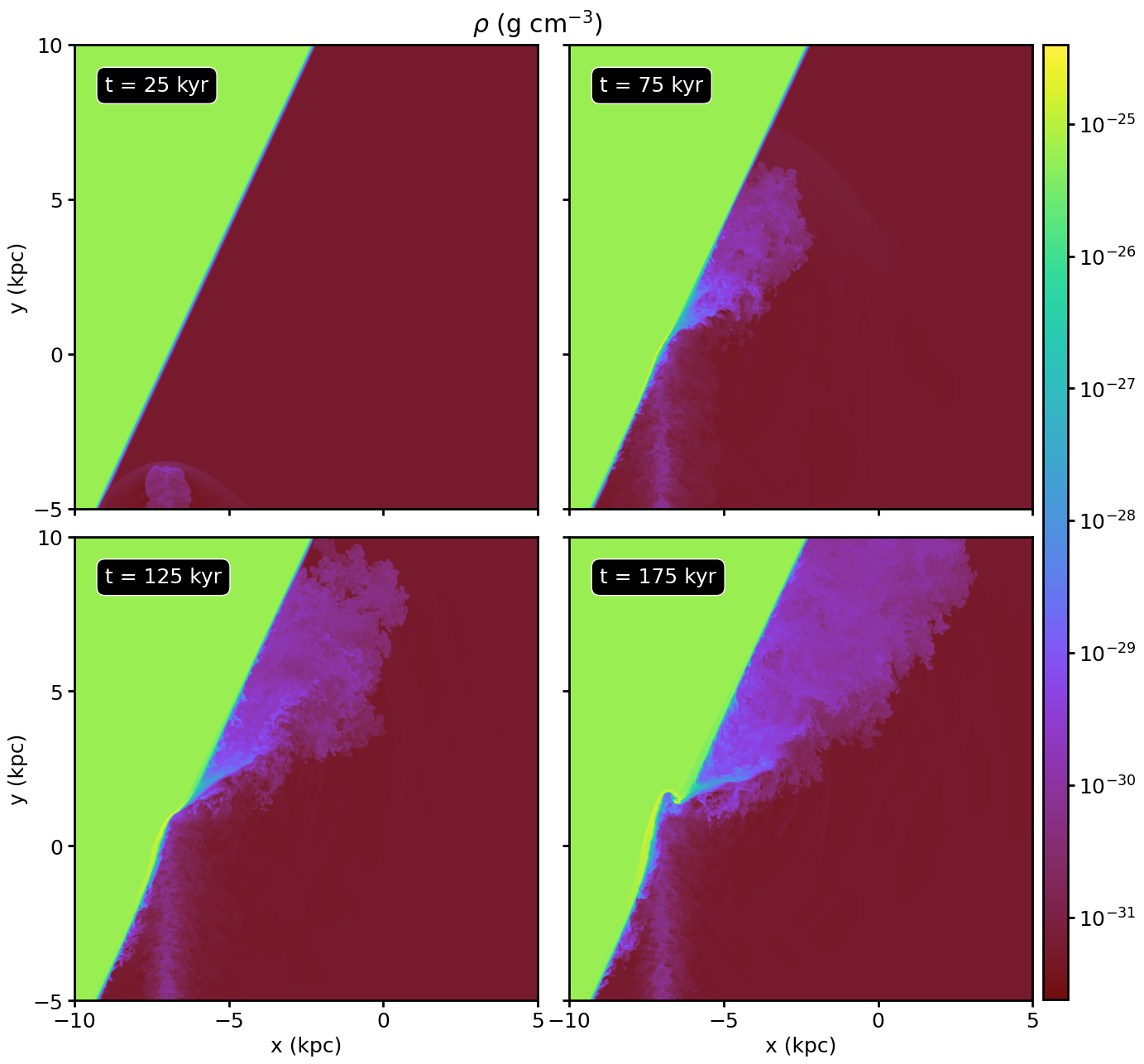}
\includegraphics[width=0.49\fulllength]{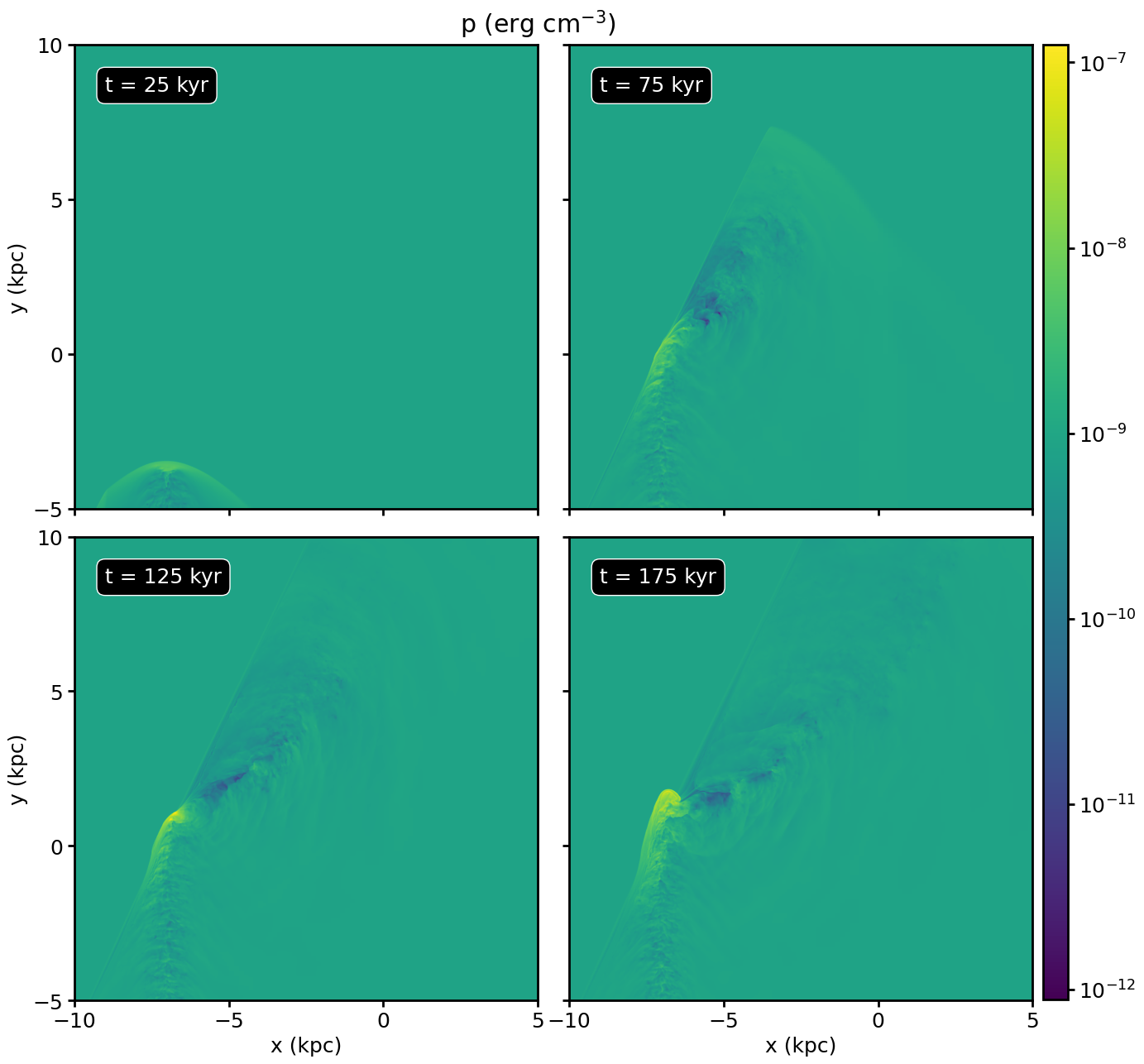}
\end{center}
\end{adjustwidth}
\caption{Density (left) and pressure (right) slices through the center of the simulation domain for the ``fiducial'' simulation, at four different epochs. Each panel is 15~kpc on a side, zoomed in slightly to focus on the parts of the simulation most affected by the jet.\label{fig:dens_pres_fid}}
\end{figure*}

\begin{figure*}
\begin{adjustwidth}{-\extralength}{0cm}
\centering
\includegraphics[width=0.49\fulllength]{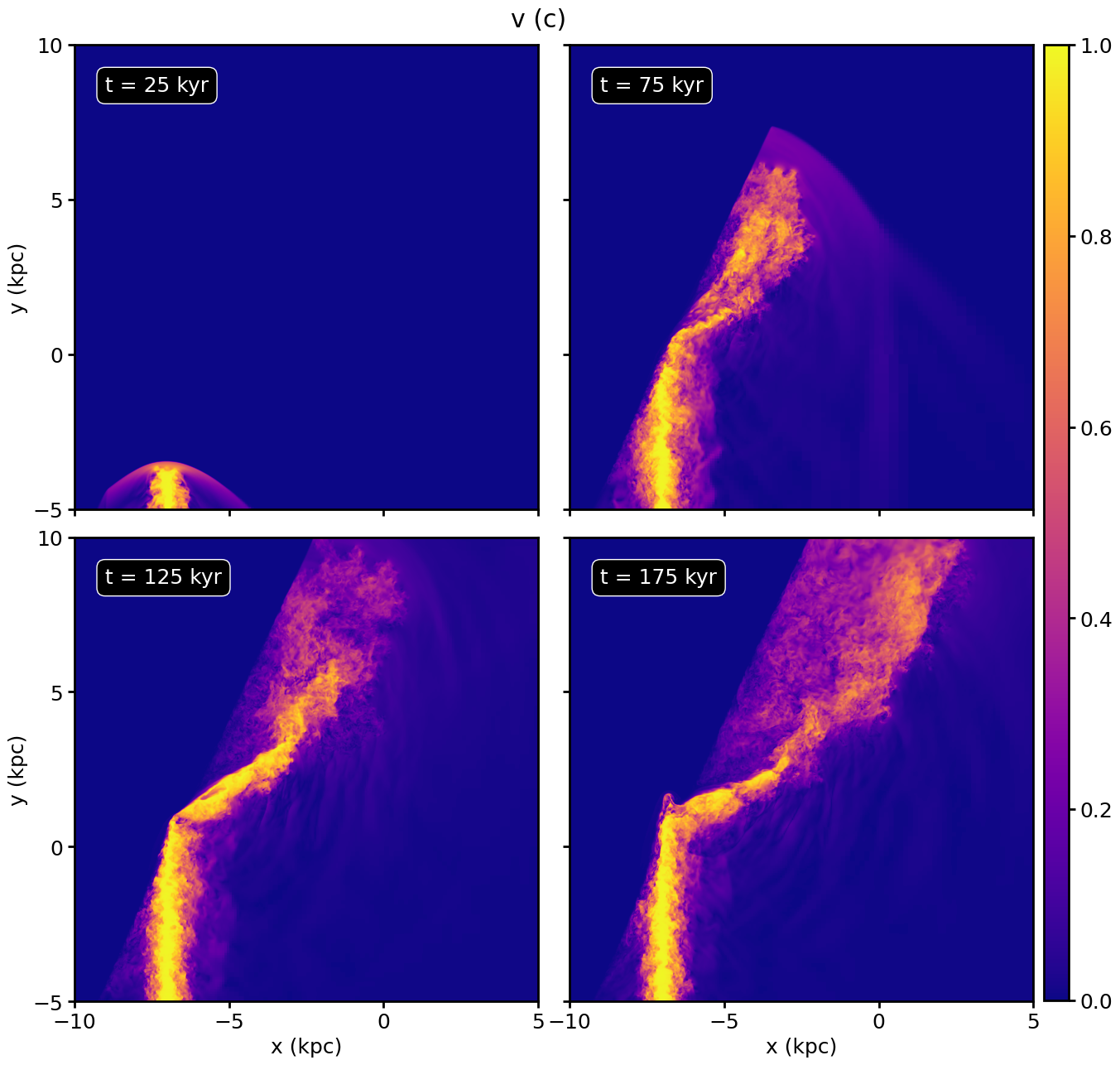}
\includegraphics[width=0.495\fulllength]{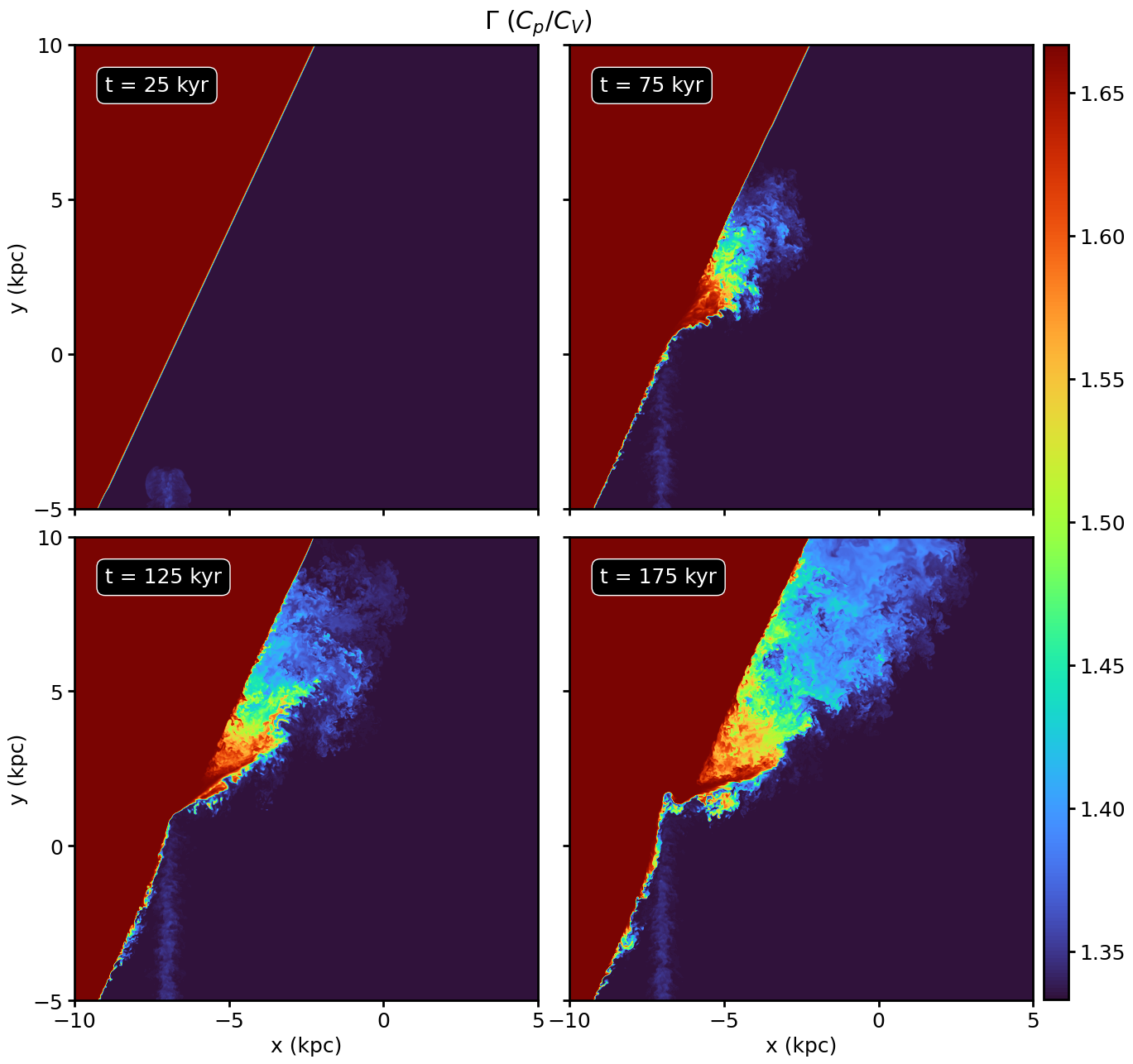}
\end{adjustwidth}
\caption{Velocity magnitude (left) and ratio of specific heats, $\Gamma$, (right) on slices through the center of the simulation domain for the ``fiducial'' simulation, at four different epochs. Each panel is 15~kpc on a side, zoomed in slightly to focus on the parts of the simulation most affected by the jet.\label{fig:velm_gamma_fid}}
\end{figure*}

We begin by describing our fiducial simulation, using the parameters detailed in
Table~\ref{tab:sim_params}. Slices through the center of the simulation domain in density and
pressure are shown in Figure~\ref{fig:dens_pres_fid}, and in velocity magnitude and ratio of
specific heats, $\Gamma$, in Figure~\ref{fig:velm_gamma_fid}. The jet approaches the interface from
the lower-left part of the domain, remaining fairly collimated at early times and driving a shock in
front of it. At approximately $t = 50$~kyr into the simulation, the jet collides with the interface.
The reflected jet continues onward, continuing to drive a shock ahead of it. The reflected jet is
highly turbulent. At the location of the collision of the jet with the interface, the gas pressure
increases by $\sim$2-3 orders of magnitude after $t \sim 100$~kyr (right panels of
Figure~\ref{fig:dens_pres_fid}), producing a ``hotspot'' where it hits the interface.

The reflected jet maintains a significant speed of $\beta\gamma \sim 1.3-2.1$ near the point of collision
with the interface, at distances larger than $\sim$5~kpc it becomes extremely turbulent and slows.
At $t \sim 125$~kyr, the incoming jet begins to burrow a hole into the ICM. As a result, the
reflected jet begins to tilt downward. As this process occurs gradually, the result is that the
outgoing material fans out into a flow directed to the right. 

\begin{figure*}
\begin{adjustwidth}{-\extralength}{0cm}
\centering
\includegraphics[width=0.49\fulllength]{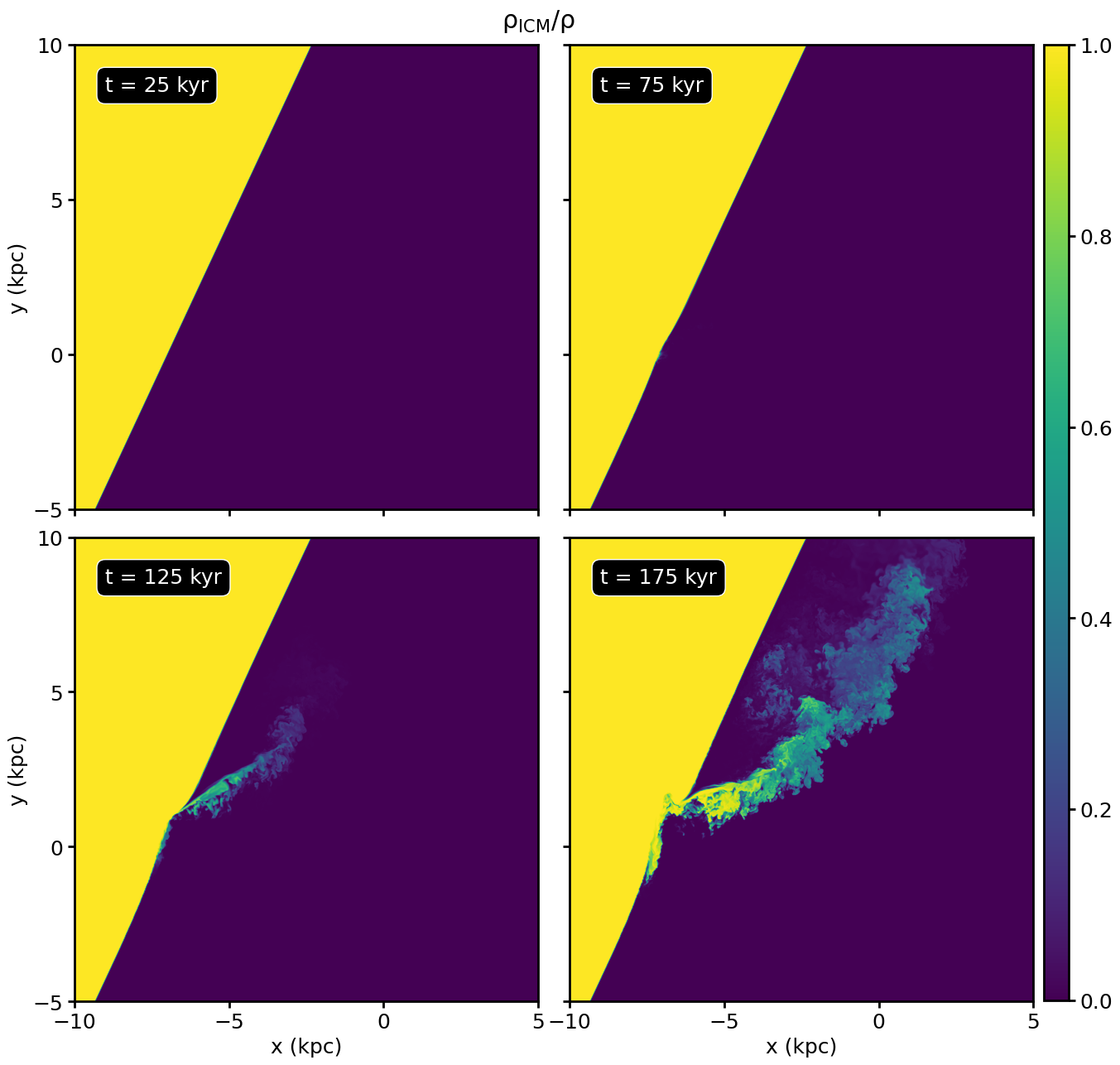}
\includegraphics[width=0.49\fulllength]{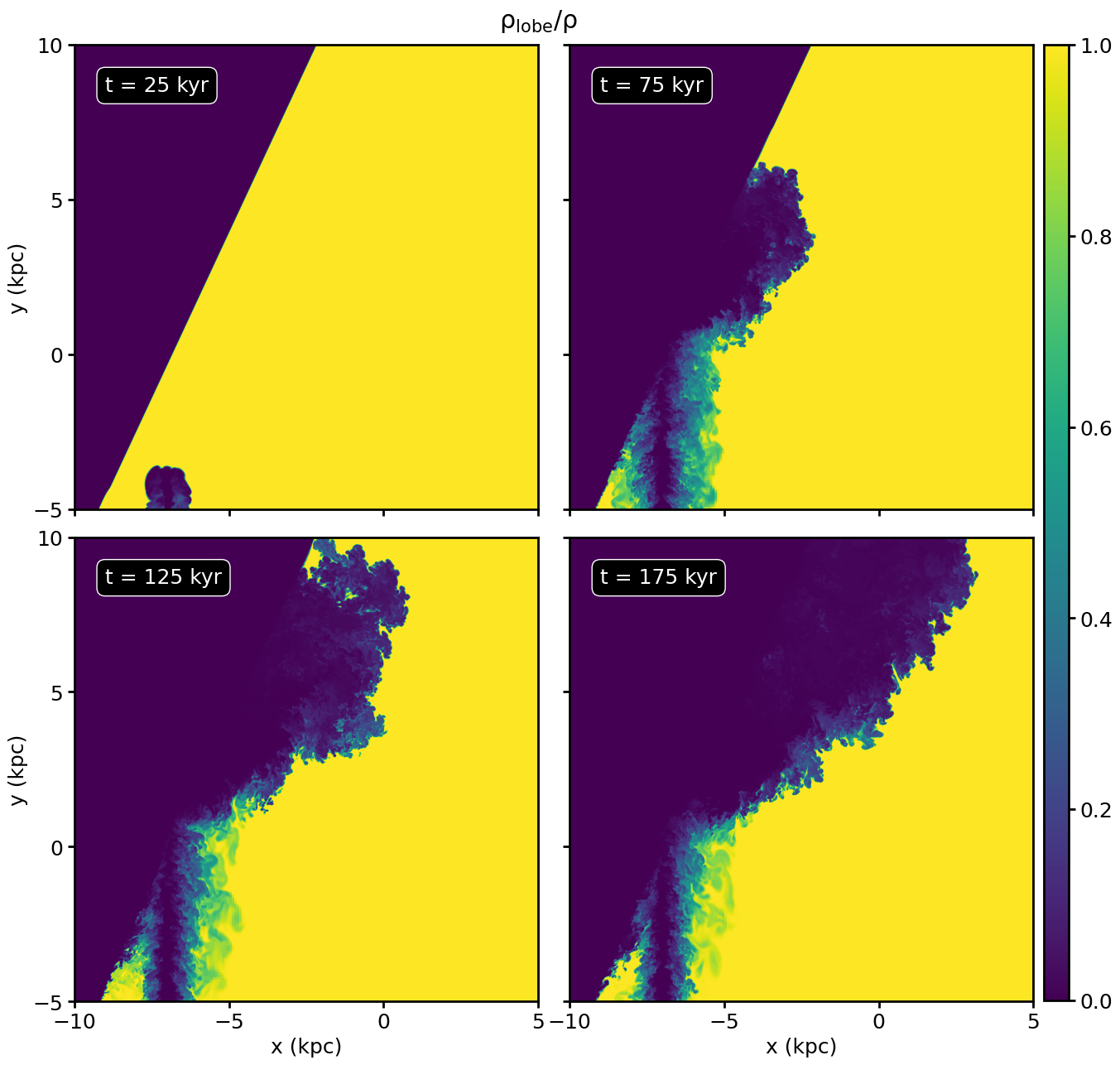}
\includegraphics[width=0.49\fulllength]{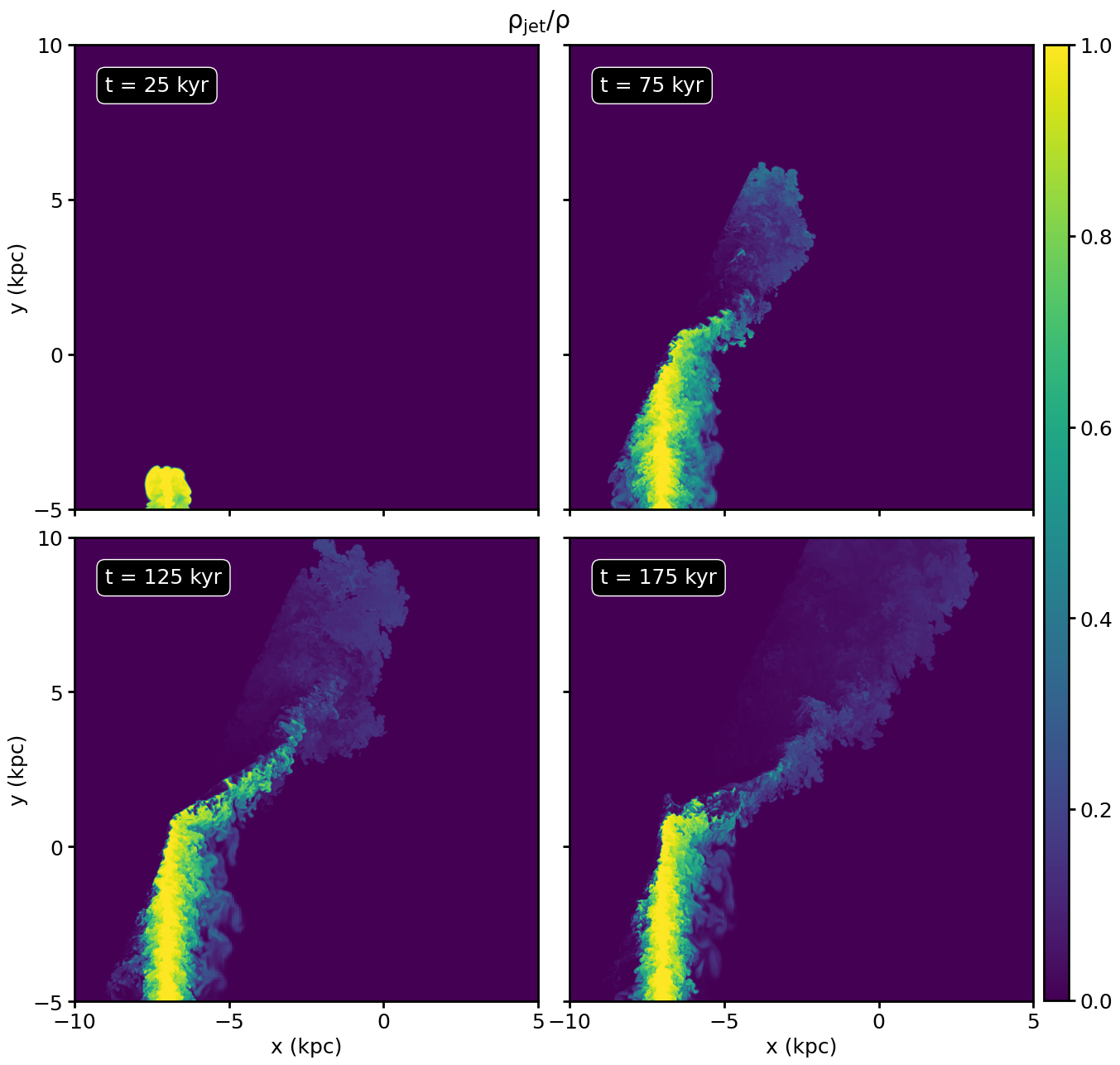}
\includegraphics[width=0.49\fulllength]{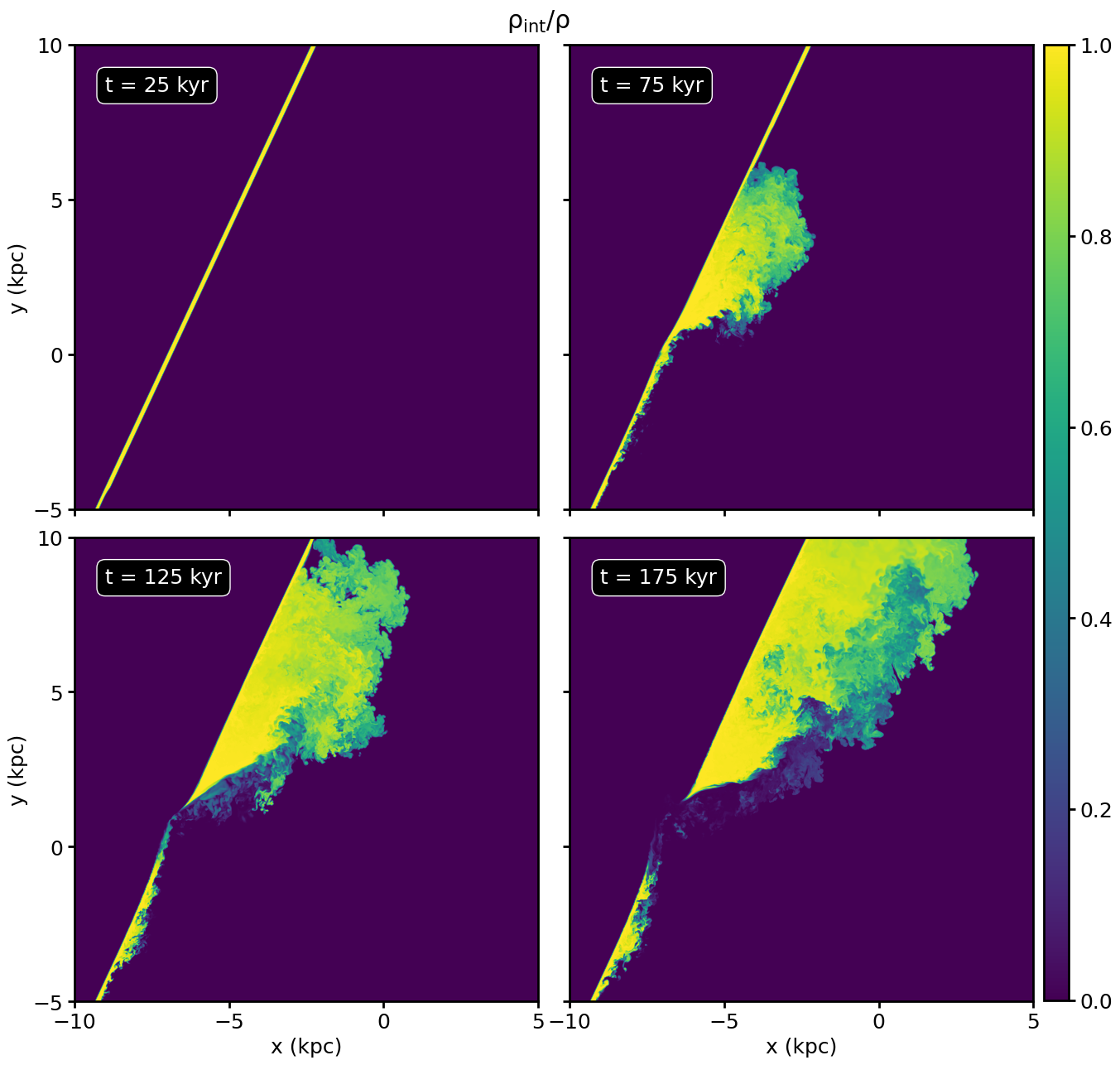}\
\end{adjustwidth}
\caption{Slices through the center of the simulation domain of the four different passive scalar fields for the ``fiducial'' simulation at four different epochs: ``ICM'' (top-left panels), ``lobe'' (top-right panels), ``jet'' (bottom-left panels), and ``interface'' (bottom-right panels) Each panel is 15~kpc on a side, zoomed in slightly to focus on the parts of the simulation most affected by the jet. \label{fig:passive_scalars}}
\end{figure*}

What is perhaps most interesting about the reflected jet flow is that very early on in its evolution
it consists mostly of material that has been stripped off of the interface. This is seen in the
right panels of Figure \ref{fig:velm_gamma_fid}, where the ratio of specific heats in much of the
outflow from the hotspot is close to 5/3, the value in the nonrelativistic ICM gas. Figure
\ref{fig:passive_scalars} shows slices of the four passive scalar fields at the same epochs plotted
in Figures \ref{fig:dens_pres_fid} and \ref{fig:velm_gamma_fid}. Initially, the reflected jet is
comprised mostly of jet material, but slowly the composition begins to include more ``interface''
gas. At $t \sim 125$~kyr, the incoming jet begins to strip material from the ICM itself and drive it
out into the lobe. The amount of lobe material in the reflected jet is negligible throughout the
simulation. As shown in the right panel of Figure \ref{fig:velm_gamma_fid}, the stripped material is
heated up as it is driven outward from the interface by thermalizing its kinetic energy or mixing with lobe plasma, and becomes nearly relativistic as the ratio of
specific heats approaches $\Gamma = 4/3$.

\subsection{Fiducial Run: Jet Power}\label{sec:jet_power}

\begin{figure*}
\begin{adjustwidth}{-\extralength}{0cm}
\centering
\includegraphics[width=0.53\fulllength]{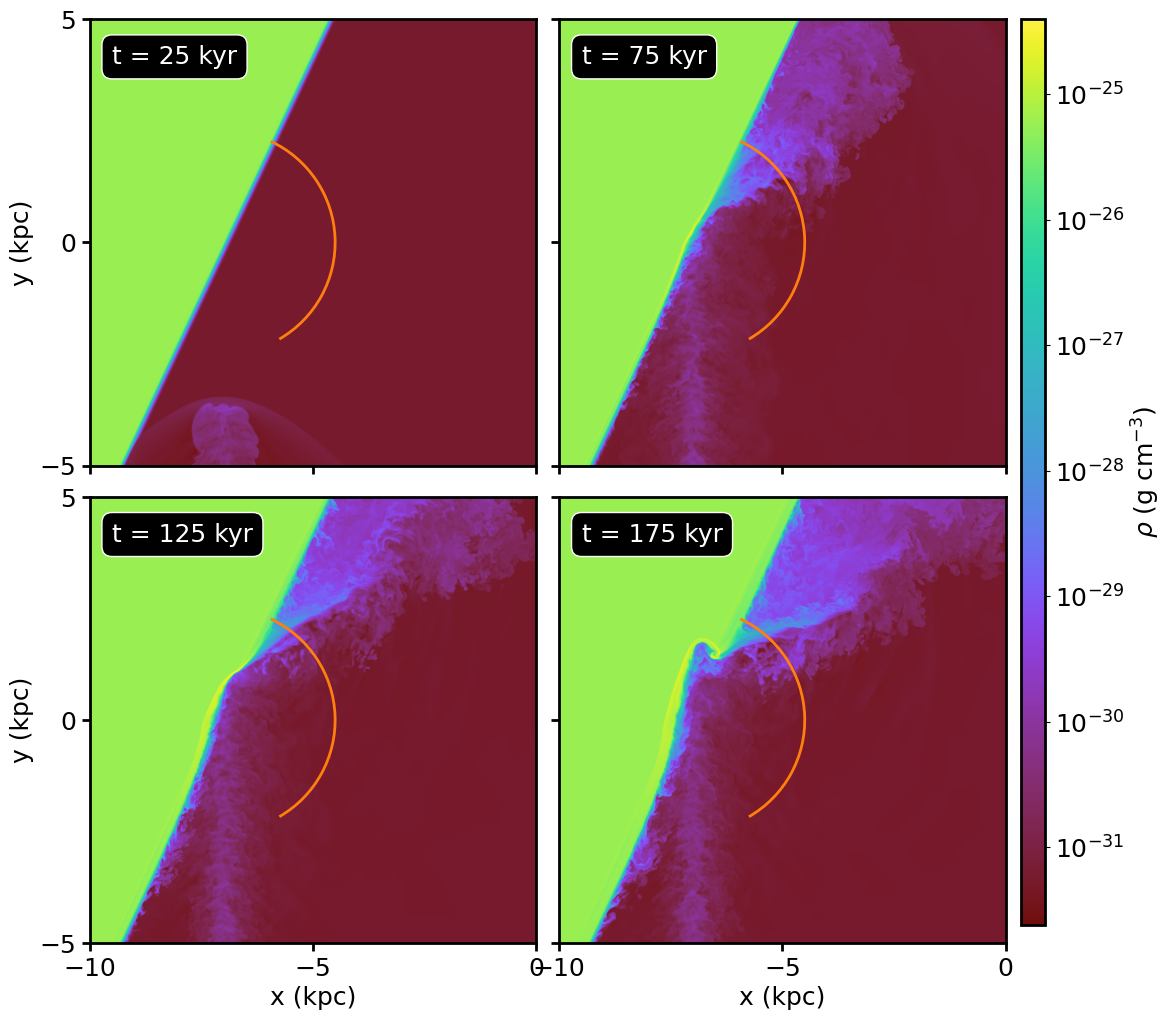}
\includegraphics[width=0.465\fulllength]{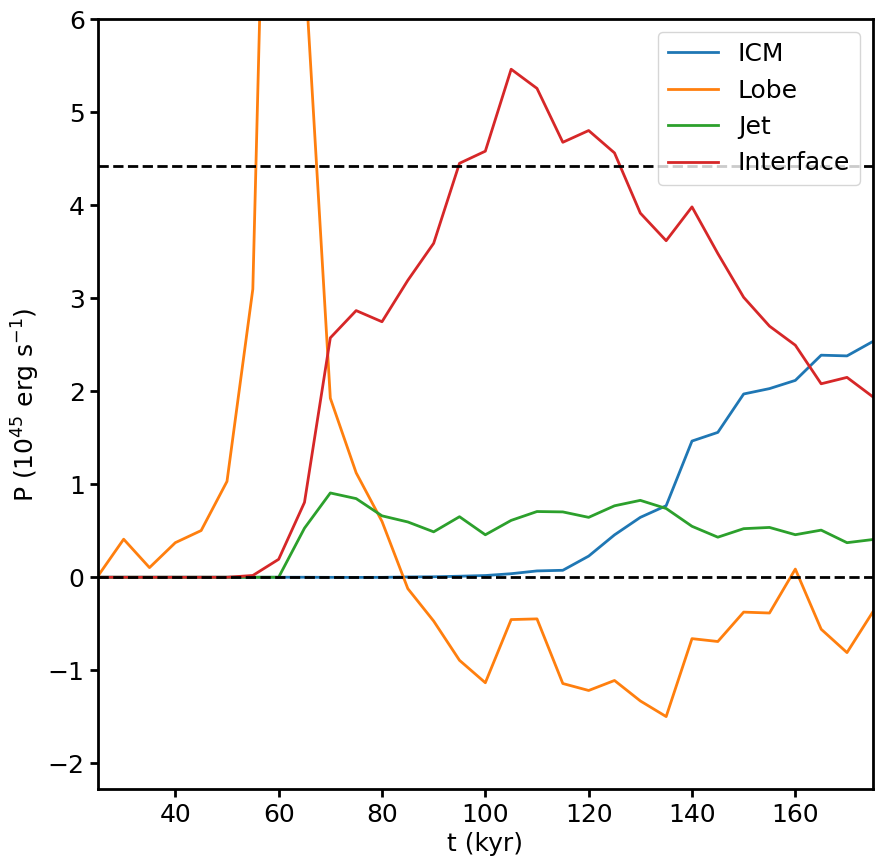}
\end{adjustwidth}
\caption{The kinetic and thermal power of the reflected jet in the ``fiducial'' simulation flowing through a portion of a spherical surface with a radius of 5~kpc, centered on the location where the jet collides with the interface. Left panels: Slices through the gas density at several epochs, showing the location of the surface in that slice in orange. Right panel: the power in the ICM, lobe, interface, and jet components flowing through the surface as a function of time.\label{fig:fluxes}}
\end{figure*}

Another way to measure the impact and composition of the reflected jet is to compute its power by integrating the flux of energy over a suitable surface. For this we use Equation \ref{eqn:jet_power} and take the surface integral over a portion of the sphere with a radius of 5~kpc, centered on the the initial point of impact of the jet on the ICM, that extends in angle from the ICM/lobe interface down to 30$^\circ$ below the $x$-axis on the negative side of the $y$-axis (see the left panels of Figure \ref{fig:fluxes} for the location of this surface drawn on slices of the gas density for several epochs). We compute the jet power associated with each of the passive scalars $\rho_c$ by integrating $P_{\rm jet,c}$ for each component separately. The resulting jet power in each component as a function of time is shown in the right panel of Figure \ref{fig:fluxes}.

The early power peak carried in the lobe gas (orange) occurs as the shock ahead of the jet crosses the surface where the power is measured.  There is also a modest peak at $\sim60\rm\ kyr$ in the power carried by jet fluid (green). However, the great bulk of the power is carried across the surface in plasma from the interface (red) from $t \sim 60-160$~kyr, and at very late times ($t \gtrsim 160$~Gyr) from the ICM (blue), both ablated from the vicinity of the hotspot. The total power flowing across this surface is comparable to the power injected via the jet (upper dashed line).

At late times, most of the kinetic power in these flows will be converted to thermal energy and the gas pressure will be approximately uniform, so that the volume occupied by each gas phase will be nearly proportional to the total energy injected with that phase. Thus, the results in Figure \ref{fig:fluxes} imply that most of the volume of the lobe will eventually be occupied by ablated interface and ICM gas.

\subsection{Varying the Width of the Interface}\label{sec:vary_width}

\begin{figure*}
\begin{adjustwidth}{-\extralength}{0cm}
\centering
\includegraphics[width=0.95\fulllength]{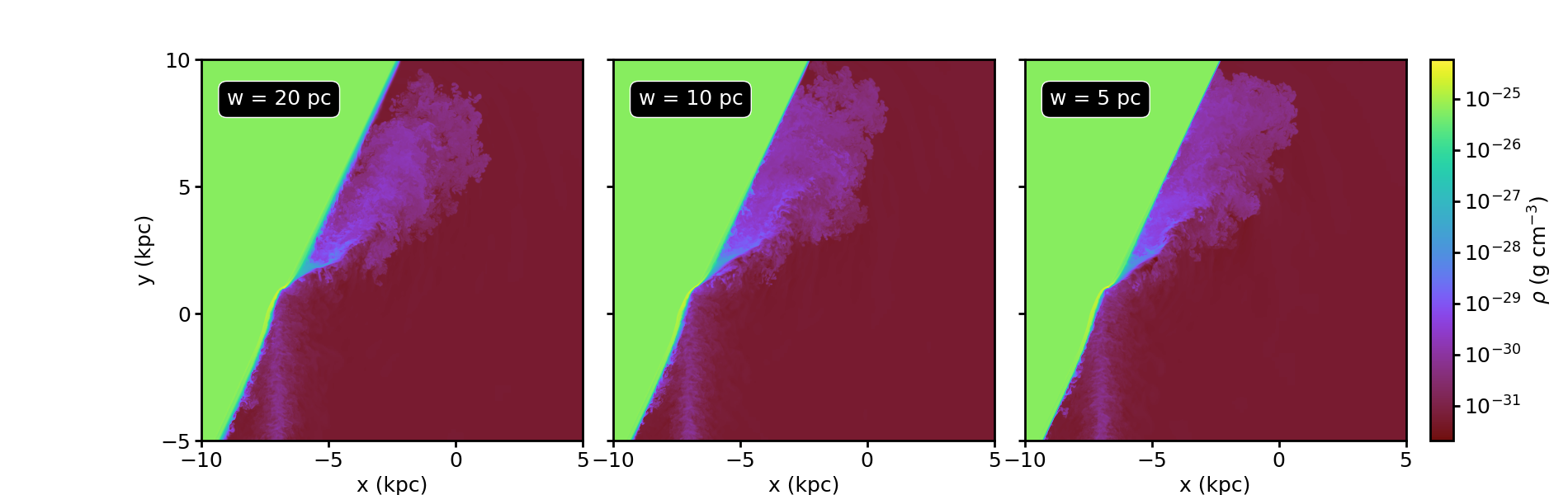}
\includegraphics[width=0.95\fulllength]{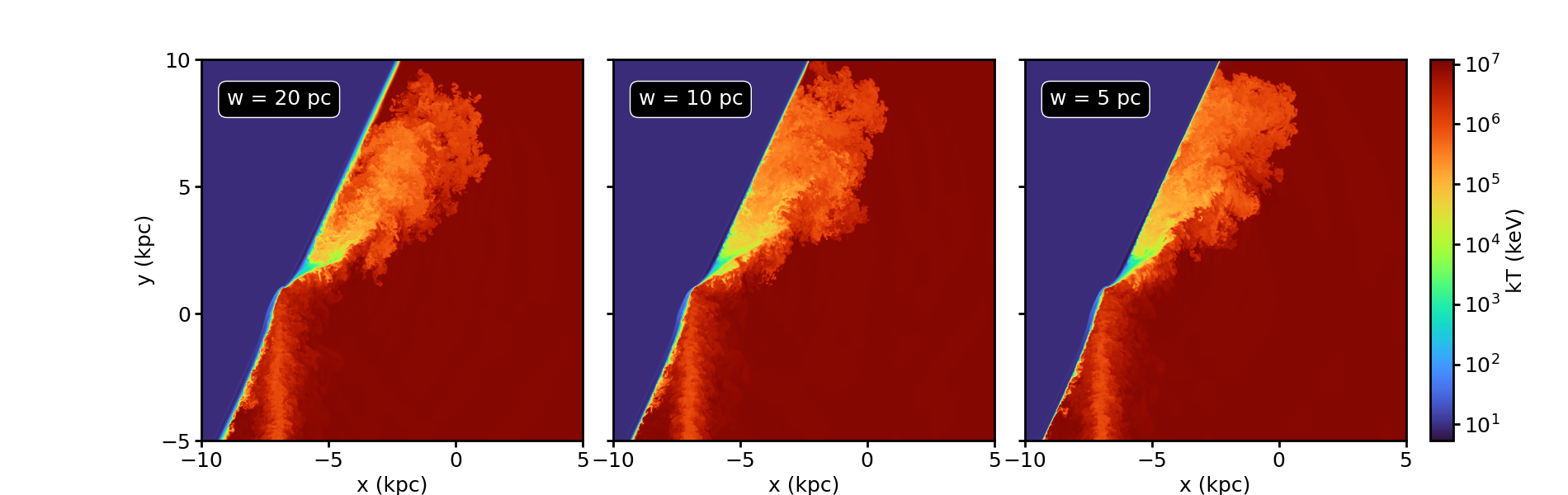}
\includegraphics[width=0.95\fulllength]{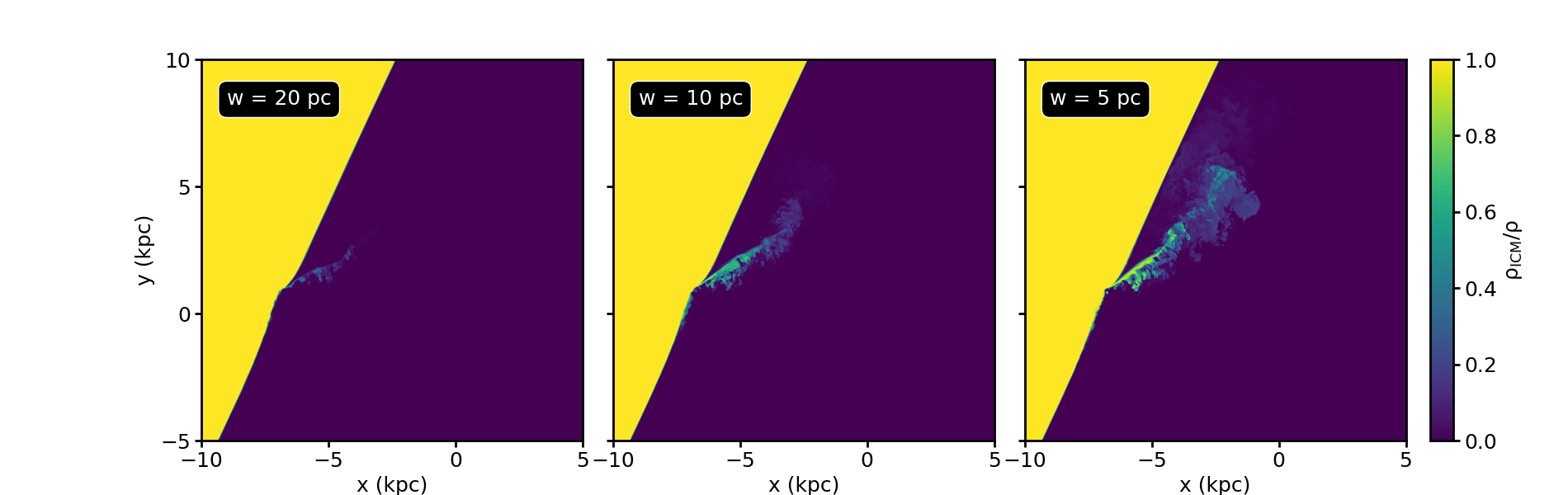}
\includegraphics[width=0.95\fulllength]{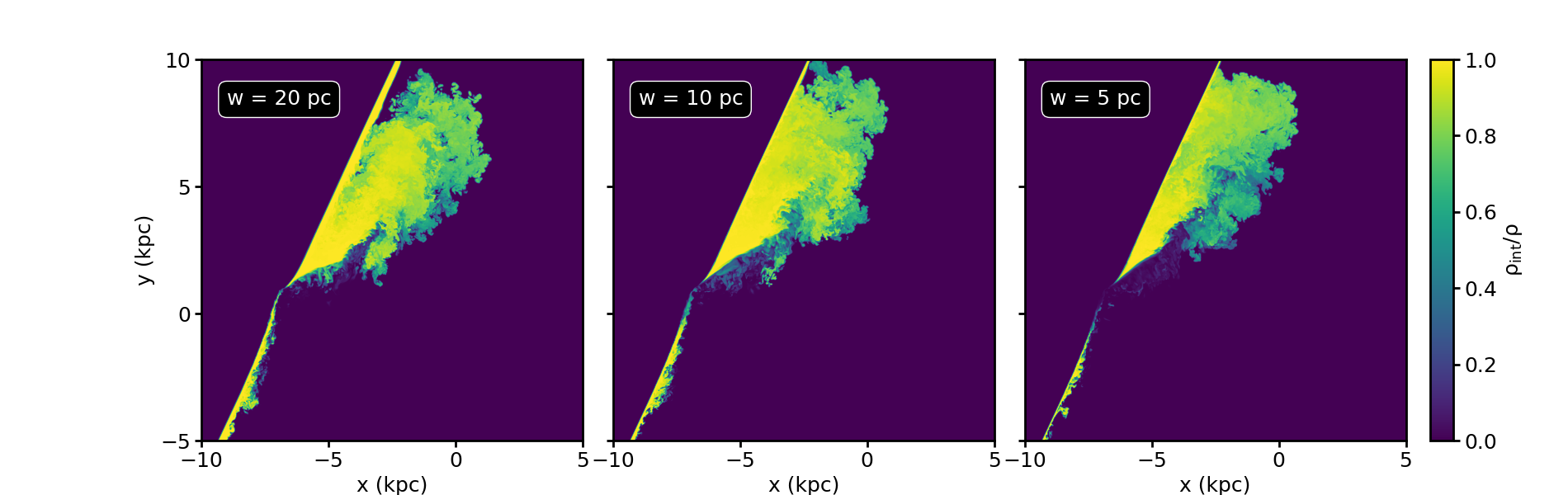}
\end{adjustwidth}
\caption{Slices at $t = 125$~kyr, showing the effect of varying the width of the interface. First row: density, second row: temperature, third row: ``ICM'' passive scalar, fourth row: ``interface'' passive scalar. Each panel is 15~kpc on a side, zoomed in slightly to focus on the parts of the simulation most affected by the jet. \label{fig:varying_width}}
\end{figure*}

\begin{figure*}
\centering
\includegraphics[width=0.5\fulllength]{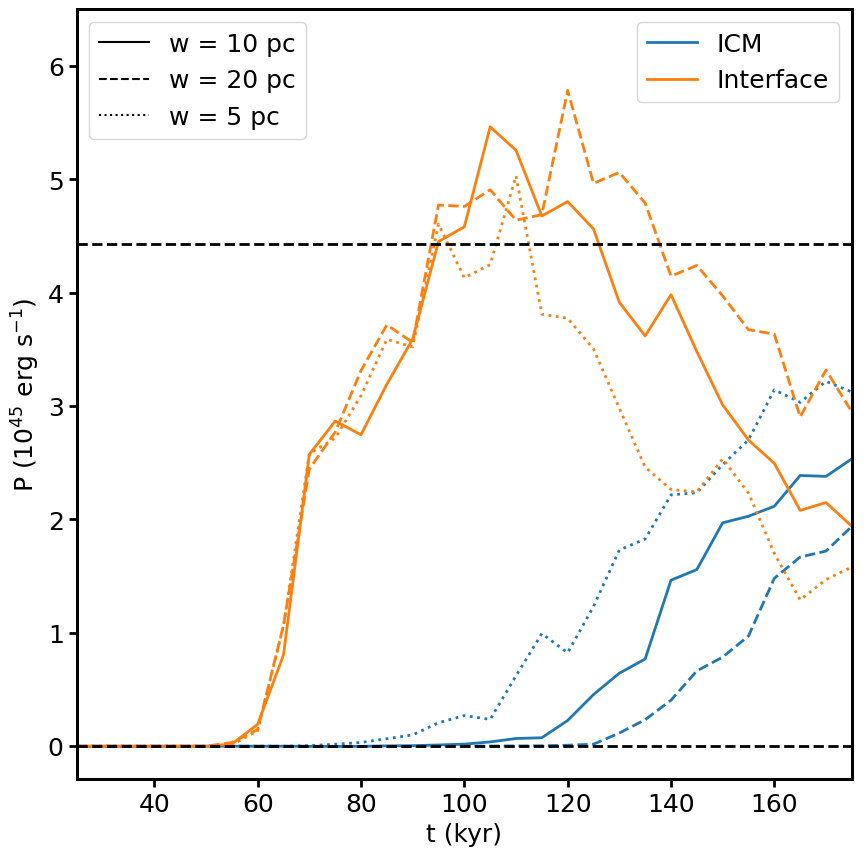}
\caption{The power in ICM, lobe, interface, and jet components flowing through the surface (shown in the left panels of Figure \ref{fig:fluxes}) as a function of time, for the simulations with varying interface width $w$.\label{fig:fluxes_diff_w}}
\end{figure*}

In this section we consider the effects of varying the width of the interface between the ICM and the lobe. Our fiducial simulation has a width parameter $w = 10$~pc; we also perform simulations with $w = 20$~pc and $w = 5$~pc. The results of this investigation are shown in Figures 
\ref{fig:varying_width} and \ref{fig:fluxes_diff_w}. In the first figure, we show slices of density, temperature, ICM, and interface material at $t = 125$~kyr. The panels show that varying the interface over this range does not affect the evolution of the reflected jet substantially--the main effect is to drive more material from the ICM off at earlier epochs if the width is smaller. This is also shown Figure \ref{fig:fluxes_diff_w}, which shows the energy fluxes of the ICM and interface material through the same surface as in Figure \ref{fig:fluxes}. The wider the interface, the later the turnover from an interface-dominated flow to an ICM-dominated one, and vice-versa.

\subsection{Carving a hole: Projected X-ray Emission}\label{sec:emission}

In this section, we consider whether Doppler beaming can account for the apparent hole around hotspot E.  As discussed in the Introduction, the energetic electrons responsible for radio synchrotron and inverse-Compton emission from the hotspot are likely accelerated to high energies in the shocks encountered there.  For a magnetic field strength of $200\ \mu\rm G$, comparable to that in hotspots A and D \citep{hcp94}, electrons with $\gamma \sim 1000$ produce synchrotron emission at $\sim1$ GHz and their synchrotron cooling times are $\sim 600,000$ yr (inverse Compton cooling times are much longer). Depending on the source of seed photons, electrons with $\gamma$ in the range $1000$ -- $10,000$ are required to produce the inverse Compton emission from the hotspots and lobe observed at $\sim 1$ keV.  At flow speeds comparable to $c$, the relativistic plasma passes through hotspot E in $\lesssim 1000$ yr, so that the radiating electrons lose little energy during their time in the hotspot. Thus, the primary source of the relativistic electrons responsible for radio and X-ray emission from the lobe is the particle acceleration that occurs in the hotspots \citep{br74}.

With a comparable population of relativistic electrons per unit volume, the plasma flowing out of hotspot E is expected to produce comparable radio and X-ray emission per unit volume to the other plasma in this vicinity.  

\begin{figure*}
\begin{adjustwidth}{-\extralength}{0cm}
\begin{center}
\includegraphics[width=0.95\fulllength]{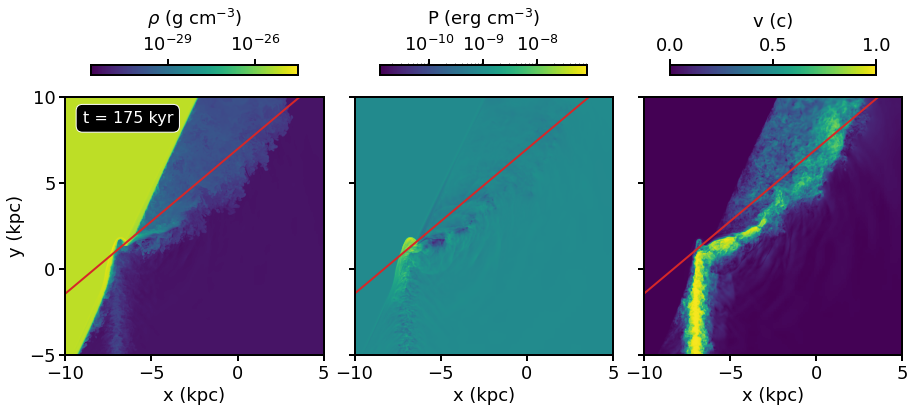}
\includegraphics[width=0.95\fulllength]{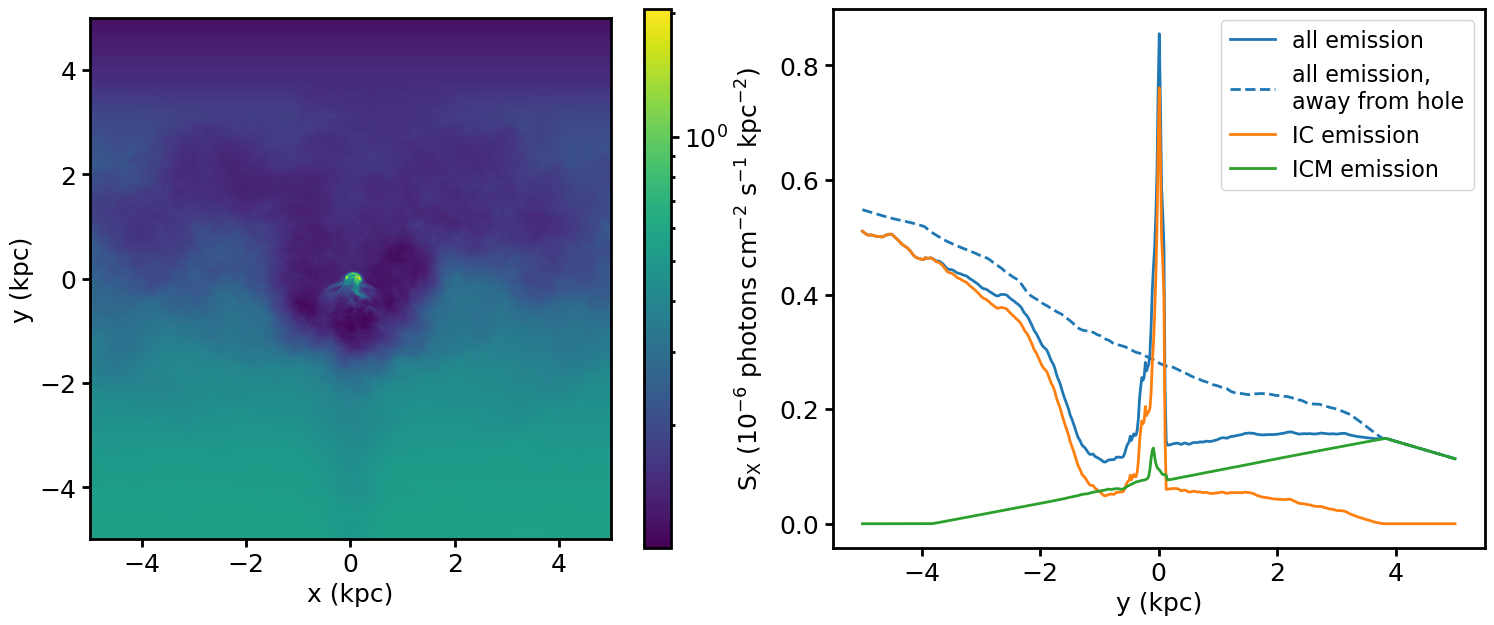}
\end{center}
\end{adjustwidth}
\caption{The appearance of a hole in X-ray emission due to Doppler de-boosting from the reflected jet. Top panels: slices through the gas density, pressure, and velocity magnitude at $t$ = 175~kyr, showing the line of sight of the projection in the bottom panels with a red line. Bottom-left panel: the projected X-ray surface brightness from all components along the chosen sight line, showing a hole of emission surrounding a hotspot. Bottom-right panel: profiles taken through the center along the $y$-axis of the map at $x$ = 0, showing separate IC-CMB and ICM thermal emission, as well as total emission in the hole region and away from the hole.\label{fig:hole}}
\end{figure*}

We model the X-ray emission from the region around the hotspot as follows: the ICM gas is modeled to have thermal emission assuming an Astrophysical Plasma Emission Code (APEC) \cite{sbl01} v3.0.9 model, where we assume the ICM has metalllicity $Z = 0.3Z_\odot$. For the relativistic plasma, we assume that it emits in the X-rays primarily via inverse-Compton scattering of cosmic microwave background photons (IC-CMB). For this component, we adopt a simple model where the emission is proportional to the energy density of the plasma for gas which satisfies $kT \geq m_ec^2$. Following \cite{dwh18}, we assume this emission is of power-law form $S_0(E) \propto E^{-\alpha}$ with $\alpha = 0.7$, and we scale the emission so that its surface brightness roughly matches that in the lobe in Cygnus A. We then adopt a sight line $\boldsymbol{\hat{\rm n}}$ which roughly intersects the hotspot and traverses through the reflected jet (top panels of Figure \ref{fig:hole}). Given the sight line, we can compute the Doppler boosting factor:
\begin{equation}
D = \frac{1}{\gamma(1-\boldsymbol{\beta}\cdot\boldsymbol{\hat{\rm n}})}
\end{equation}
and multiply the factor $D^{3-\alpha}$ \cite{Lind1985} by the emission at every point. We compute the total emission of both components in the 0.5-7.0~keV band and project along the chosen sight line. 

The total projected emission in the vicinity of the pressure hotspot along the line of sight is shown in the bottom-left panel of Figure \ref{fig:hole}. The bright hotspot appears in the center of the image, surrounded by a cavity in emission roughly $\sim4$~kpc in diameter, produced by the Doppler de-boosting of the reflected jet along the sight line. The bottom-right panel of Figure \ref{fig:hole} shows profiles taken along the $y$-axis of the map at $x$ = 0, showing the total emission (blue) along this profile and the IC (orange) and ICM (green) emission separately. A dashed blue line shows the total emission profile at another location on the $x$-axis, far away from the hotspot and the hole which surrounds it, showing the hole creates a significant deficit of emission. The resulting hole is roughly a factor of two smaller in diameter than the observed feature in Cygnus A, but the simulation qualitatively reproduces the general features.  Note that the overall decrease in brightness with increasing $y$ is partly due to the decreasing lengths of the sight lines within the lobe, an artifact of the rectangular simulation box.

\section{Discussion \& Conclusions}\label{sec:summary}

The nearby FRII radio galaxy Cygnus A exhibits hotspots of emission at the point of interaction of its jets with the interface between the expanding lobe plasma which surrounds the jets and the ICM, as is typical for such sources. In deep \textit{Chandra} X-ray observations of Cygnus A, \cite{sjn20} identified a ``hole'' in the emission surrounding one of the hotspots, and suggested that it was produced by Doppler de-boosting of a reflected jet moving away from the hotspot and from the observer along the line of sight. 

In this work, we have presented relativistic hydrodynamic simulations of a FRII jet impinging on the interface between a relativistic plasma in the lobe surrounding the jet and the ICM, using a simple plane-parallel model for the interface with the axis of the jet inclined to the interface. The interaction of the highly collimated jet with the interface produces a high-pressure ``hotspot'' and a turbulent reflected jet that fans out into the lobe and fills it with somewhat denser gas, first from the interface transition region and at later times from the ICM itself. The amount of material in the reflected jet from the incident jet itself is always negligible except immediately after the initial collision, as the low-density plasma from the jet mixes with the much more dense material in the interface and ICM.  This indicates that interaction of the jet with the ICM will fill the lobe with relatively energetic ions originating from the ICM, as proposed by \cite{br74}. This general behavior is not affected significantly by varying the width of the interface. If viewed along a line of sight approximately aligned with the reflected jet, with the latter moving away from the observer, the emission from this gas surrounding the hotspot is Doppler de-boosted, producing a cavity or ``hole'' of emission surrounding the hotspot, as seen in Cygnus A. 

This simple scenario qualitatively reproduces the essential features of the hotspot and its surrounding hole. Future work will include a more accurate representation of the gas physics, including magnetic fields, which are expected to be dynamically significant in and near the hotspot, as well as acceleration of relativistic particles by internal shocks in the jet. The inclusion of magnetic fields and a more accurate model for particle acceleration will also permit more sophisticated models of the radio synchrotron and X-ray synchrotron-self-Compton emission from the hotspot and surrounding region. The flow onward from hotspot E does not appear sufficiently well collimated to produce hotspot D \cite{sjn20}, although the collimation will be altered by the inclusion of magnetic fields. A model including magnetic fields, along with a curved interface to allow the production of multiple hotspots, will be required to test this hypothesis.

Models of relativistic jets interacting with inclined or curved lobe/ICM interfaces may also be applied to other observed interactions, such as the eastern jet of Cygnus A and the FRII source 3C 220.1, which shows a jet potentially deflected against an interface at multiple locations \cite{Liu2020}.  

\vspace{6pt} 

\authorcontributions{Conceptualization, P.N. and J.Z.; methodology, P.N. and J.Z.; software, J.Z., P.-H.T., and H.-Y.S.; validation, J.Z.; formal analysis, J.Z. and P.N.; investigation, J.Z. and P.N.; resources, J.Z.; data curation, J.Z.; writing---original draft preparation, J.Z. and P.N.; writing---review and editing, J.Z., P.N., P.-H.T., H.-Y.S., and T.J.; visualization, J.Z.; funding acquisition, J.Z. All authors have read and agreed to the published version of the manuscript.}

\funding{JAZ and PEJN are funded by the Chandra X-ray Center, which is operated by the Smithsonian Astrophysical Observatory for and on behalf of NASA under contract NAS8-03060. P-ST and H-YS are supported by the National Science and Technology Council (NSTC) of Taiwan under Grants No. NSTC 111-2628-M-002-005-MY4 and No. NSTC 108-2112-M-002-023-MY3. TWJ is supported by NSF grant AST2205885 to the University of Minnesota.}

\dataavailability{Simulation data will be made available by reasonable request to the authors.} 

\acknowledgments{The simulations were run on the Pleiades supercomputer at NASA/Ames Research Center. Software packages used in this work include: GAMER\footnote{\url{https://github.com/gamer-project/gamer}} \cite{Schive2010,Schive2018,Tseng2021}; AstroPy\footnote{\url{https://www.astropy.org}} \cite{AstroPy2013};
Matplotlib\footnote{\url{https://matplotlib.org}} \cite{Hunter2007};
NumPy\footnote{\url{https://www.numpy.org}} \cite{Harris2020};
yt\footnote{\url{https://yt-project.org}} \cite{Turk2011}}

\conflictsofinterest{The authors declare no conflict of interest.} 

\appendixtitles{yes} 
\appendixstart
\appendix
\section[\appendixname~\thesection]{Resolution Test}

\begin{figure*}
\begin{adjustwidth}{-\extralength}{0cm}
\begin{center}
\includegraphics[width=0.49\fulllength]{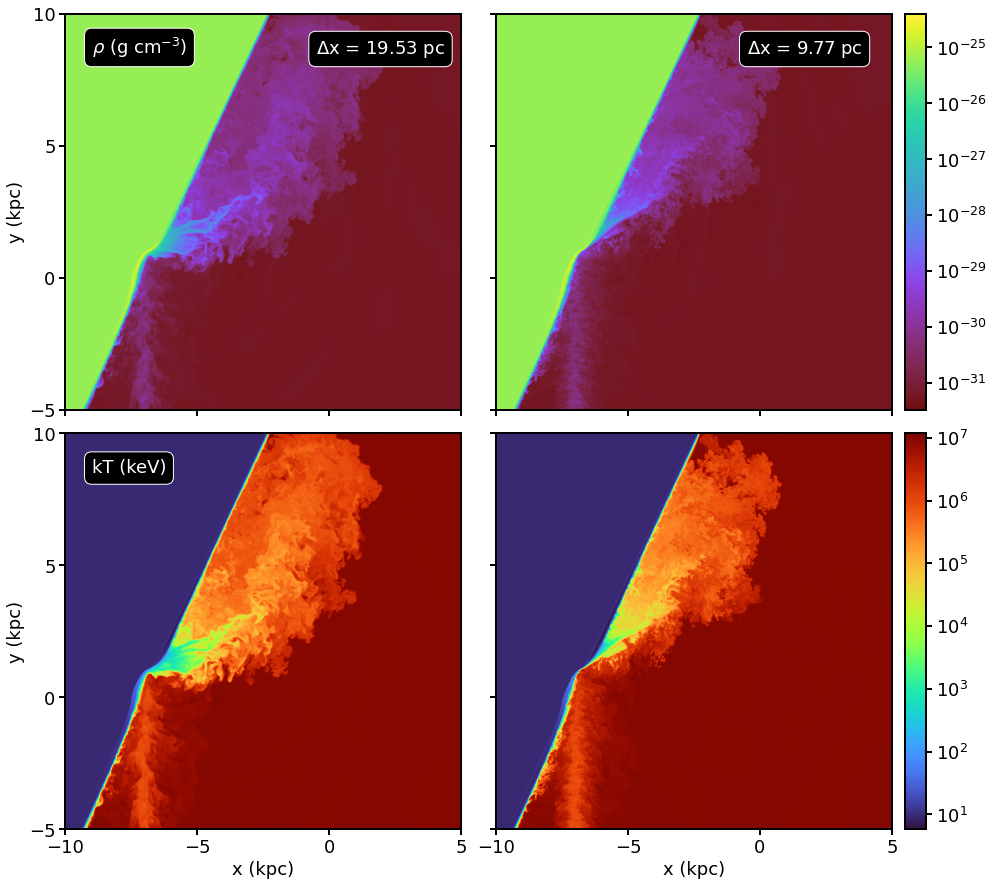}
\includegraphics[width=0.49\fulllength]{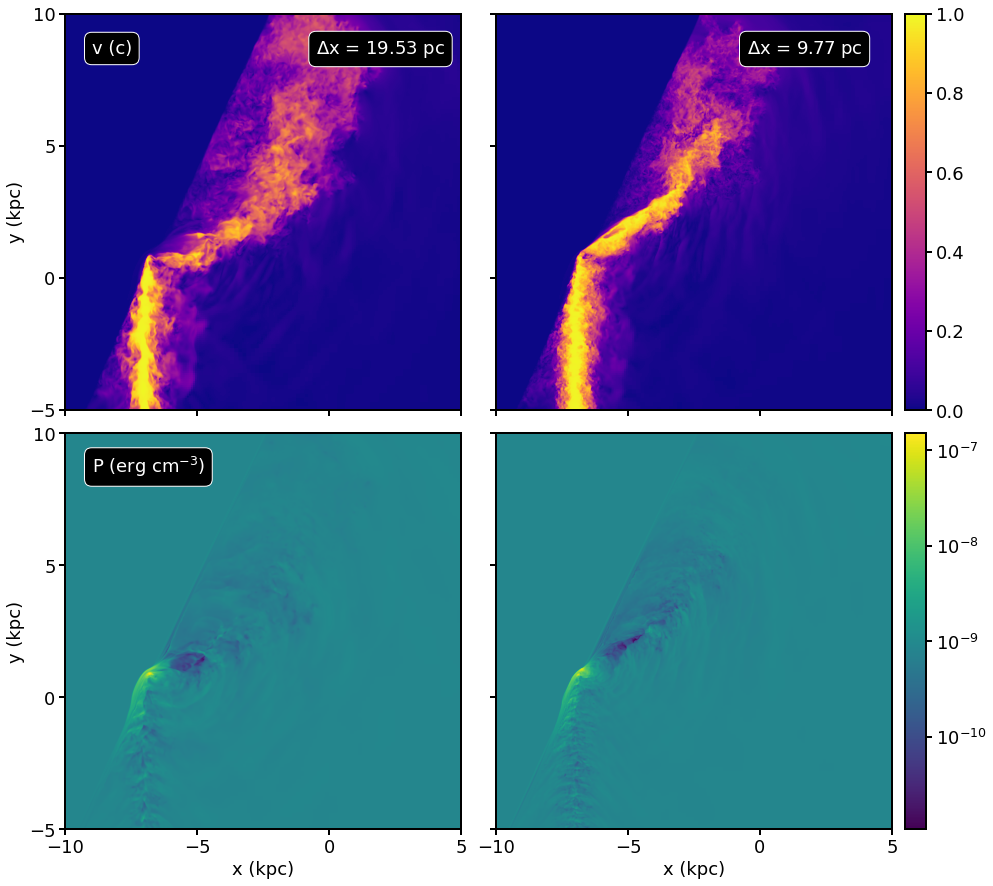}
\end{center}
\end{adjustwidth}
\caption{Comparison of the fiducial simulation at two different spatial resolutions at $t = 125$~kyr. Left panels: Slices through the center of the simulation domain of density and temperature. Right panels: Slices of velocity magnitude and pressure. Each panel is 15~kpc on a side, zoomed in slightly to focus on the parts of the simulation most affected by the jet.\label{fig:dens_T_v_p_res}}
\end{figure*}

\begin{figure*}
\begin{adjustwidth}{-\extralength}{0cm}
\begin{center}
\includegraphics[width=0.49\fulllength]{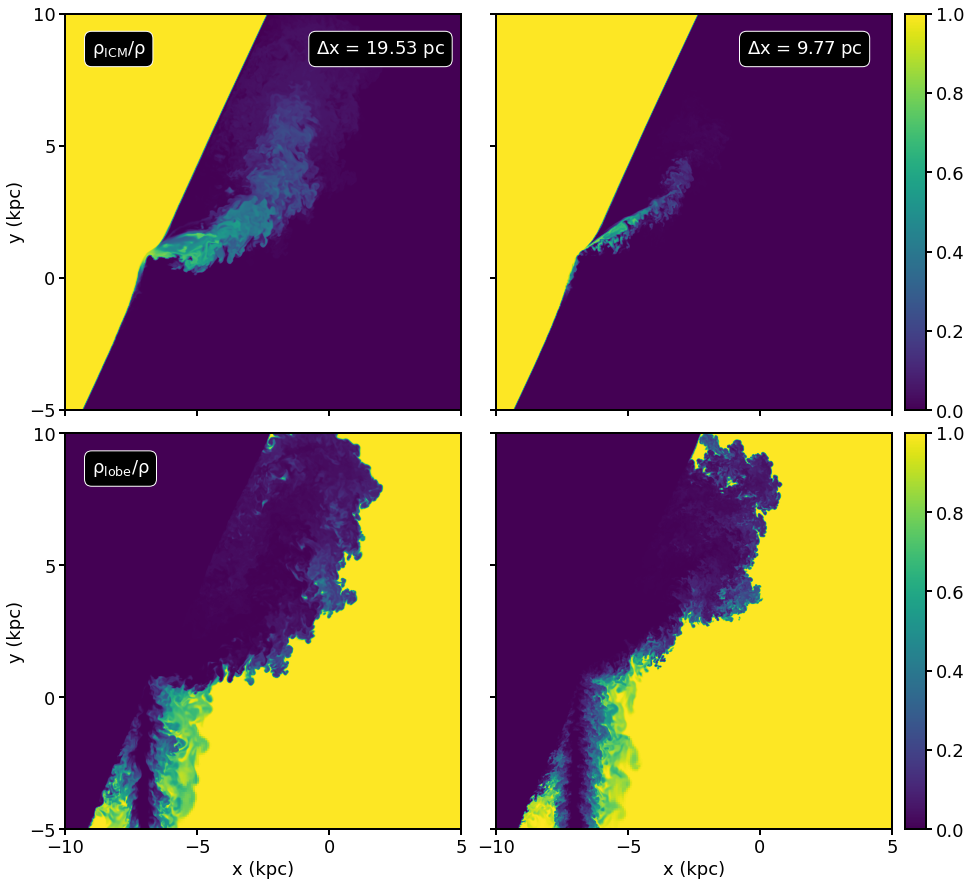}
\includegraphics[width=0.49\fulllength]{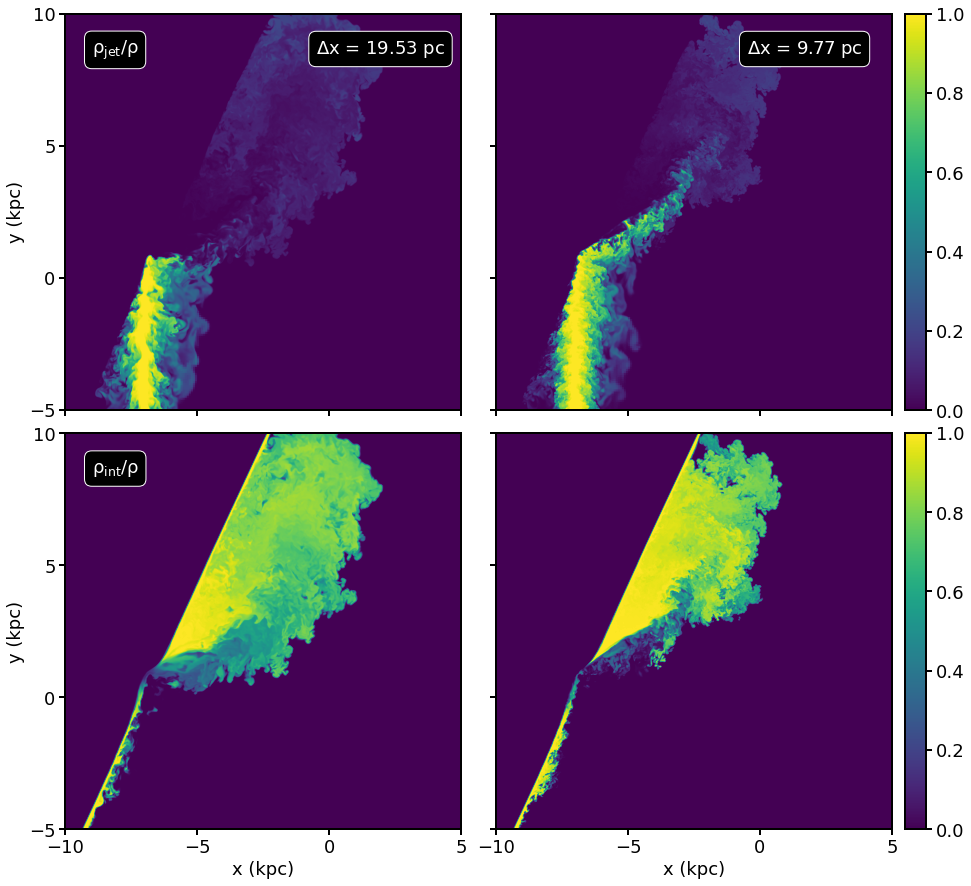}
\end{center}
\end{adjustwidth}
\caption{Comparison of the fiducial simulation at two different spatial resolutions at $t = 125$~kyr. Panels show slices through the center of the simulation domain of the four different mass scalars. Each panel is 15~kpc on a side, zoomed in slightly to focus on the parts of the simulation most affected by the jet. \label{fig:scalars_res}}
\end{figure*}

In this Appendix we present a comparison between our fiducial simulation with a finest cell width of $\Delta{x} \approx 10$~pc and an otherwise identical simulation with a finest cell width of $\Delta{x} \approx 20$~pc (one less level of refinement). With coarser spatial resolution, features that vary on short length scales such as the narrow interface and the velocity profile of the jet are most affected. Plasma of different entropies and origins (mass scalars) is also mixed on coarser length scales. Slices of various quantities at $t = 125$~kyr in the two simulations are shown in Figures \ref{fig:dens_T_v_p_res} and \ref{fig:scalars_res} to illustrate these effects. 

Qualitatively, the main features of the fiducial simulation are reproduced in the coarser-resolution version--a turbulent reflected jet that consists initially mostly of material stripped from the interface, and at later times incorporates more material from the ICM itself. The reflected jet is somewhat more collimated near the interface in the higher-resolution simulation, and contains more jet material, indicating that the mixing of the latter with the interface material is more complete at lower resolution.

\begin{adjustwidth}{-\extralength}{0cm}

\reftitle{References}


\bibliography{ms.bib}

\PublishersNote{}
\end{adjustwidth}
\end{document}